# Advancing characterisation with statistics from correlative electron diffraction and X-ray spectroscopy, in the scanning electron microscope

T.P. McAuliffe*, A. Foden, C. Bilsland, D. Daskalaki-Mountanou, D. Dye & T.B. Britton

Department of Materials, Prince Consort Road, Imperial College London, London, SW7 2AZ

*t.mcauliffe17@imperial.ac.uk

## Abstract

The routine and unique determination of minor phases in microstructures is critical to materials science. In metallurgy alone, applications include alloy and process development and the understanding of degradation in service. We develop a correlative method, exploring superalloy microstructures which are examined in the scanning electron microscope (SEM) using simultaneous energy dispersive X-ray spectroscopy (EDS) and electron backscatter diffraction (EBSD). This is performed at an appropriate length scale for characterisation of carbide phases' shape, size, location, and distribution. EDS and EBSD data are generated using two different physical processes, but each provide a signature of the material interacting with the incoming electron beam. Recent advances in post-processing, driven by 'big data' approaches, include use of principal component analysis (PCA). Components are subsequently characterised to assign labels to a mapped region. To provide physically meaningful signals, the principal components may be rotated to control the distribution of variance. In this work, we develop this method further through a weighted PCA approach. We use the EDS and EBSD signals concurrently, thereby labelling each region using both EDS (chemistry) and EBSD (crystal structure) information. This provides a new method of amplifying signal-to-noise for very small phases in mapped regions, especially where the EDS or EBSD signal is not unique enough alone for classification.

## 1. Introduction

Progress towards rapid, accurate and statistically robust characterisation of microstructures has been made in recent years with developments in both experimental techniques and data processing. There has been interest in 'correlative' microscopy, where multiple techniques are employed to access independent information channels sampled from the same area of interest (AOI) [1–3]. Successful use of correlative microscopy yields superior characterisation capability (as limitations of individual techniques may be mitigated) and provides better confidence in phase assignment if independent classifications are mutually inclusive. We develop correlative microscopy through combining electron backscatter diffraction (EBSD) and energy dispersive X-ray spectroscopy (EDS), with maps collected simultaneously in a scanning electron microscope (SEM). The concept and approach is applicable to other techniques in which a measurement (of a spectrum, diffraction pattern, *etc*) is made at one of many known scan locations, such as in scanning transmission microscopy or 4D-STEM [4].

Confidently assessing microstructure is of significant concern in materials science and engineering, as well as in the earth and planetary sciences. In the present work, we develop a new approach using an example in Co/Ni-base superalloys. In these alloys, there are carbide precipitates that are known to strongly influence fatigue and tertiary creep performance [5–9]. The precipitates are thought to increase boundary cohesivity and to mitigate sliding. However, their high temperature oxidation reduces grain boundary strength and permits easier intergranular crack propagation. Some precipitate phases are thought to exhibit better oxidation properties than others, conferring superior enviro-mechanical stability across deformation regimes [10]. To assist in understanding these phases, we can use EBSD and EDS analysis for characterisation. With EDS alone, it can be difficult to distinguish two phases of similar chemistry but different structure, for example $M_{23}C_6$ and $M_6C$ carbides. Similarly, using EBSD alone it can be difficult to distinguish two phases of similar structure but different chemistry, for example the pseudo-FCC matrix and MC carbide. Applying correlative EBSD and EDS offers a solution to this problem.

Very briefly, the 2D EBSD pattern captured using conventional EBSD is created from near surface (<20 nm) scattering and diffraction events [11]. The raw signal within the EBSD pattern is semi-quantitative, due to the many transfer processes and image processing stages required to generate useful patterns for analysis. These patterns can be indexed to reveal the orientation and structure of the crystal, provided the signal-to-noise ratio is high enough and a series of input phases are used as input 'classifiers' which the indexing algorithm is tested against. EBSD analysis is challenging if two phases have similar crystal structures (e.g. only slight changes in lattice parameter or subtle differences between symmetrically-related structures, particularly if an orientation relationship is present) or the signal strength is poor (e.g. a small phase). Signal classification can be improved through the use of template matching [12–16] against simulated patterns using dynamical diffraction theory [11,17–19].

To amplify signal-to-noise for poor quality patterns, principal component analysis (PCA) can be used. PCA is a data processing approach used to reduce and project measured variables of a group of objects onto an orthogonal set of basis vectors: the principal components. The PCA method is described in detail elsewhere [20], [21]. This approach is useful, as it can amplify signal-to-noise in data, especially where the data at each point is oversampled and noisy (e.g. a diffraction pattern or a highly resolved energy dispersive X-ray spectrum) and these measurements can be used as the PCA 'variables'. The dimensions of these variables are the diffraction vectors (the intensities of each pixel in a vectorised EBSD pattern) and the counts for each EDS spectrum energy bin. When a scan is performed on a sample in a scanning electron microscope one 'object' (a full set of measurements) is collected per measurement 'point'.

For the present work, we can apply PCA, but we note that the principal components of diffraction data may be difficult to interpret. This is because PCA extracts and ranks principal component vectors by the strength of each component signal, and many signals will contribute to each point in our map. From the physics of our problem, we know (broadly) that the variance of the signal between one phase and another should be similar, and that we would like (typically) only one signal type to label each point in the map. This motivates us to develop the work of Wilkinson *et al* [13] and Brewer *et al* [22] who have shown that a rotation of the set of principal component EBSD patterns that maximises the variance between each member of the set effectively reduces a full EBSD dataset down to a single representative, or 'characteristic' EBSD pattern for each commonly labelled domain. If the number of components is well selected, prior to VARIMAX rotation, then these can correspond to a single pattern per grain, and for oversampled or deformed grains the domains may also correlate with sub-grains. In this work, we refer to the VARIMAX rotated components as 'rotated characteristic components' (RCCs). Each RCC contains a characteristic electron backscatter diffraction pattern (RC-EBSP).

In our work, we use the Wilkinson *et al* [13] method as our starting point, and now address the challenge of including EDS spectra. Each EDS spectrum contains chemical information related to the interaction volume associated with the generation and escape of X-rays, which are counted by a detector. The number of X-rays generated for each energy are a function of the electron transitions. Characteristic X-rays are generated from the primary beam promoting a core electron, and that core electron subsequently 'falling down' to a lower energy level to generate the radiation. These peaks are superposed on the Bremsstrahlung. The spectra contain digitised signals of the number of counts per energy level, as detected (in our case) using a silicon drift detector (SDD). The signal also contains a broadening function related to the detector and instrument noise [23].

In the first instance we can append the EDS spectra onto the end of the diffraction pattern vector. However, in practice the variance in the EDS and EBSD signals may be significantly different, and the number of channels in each can vary significantly. These properties are important for our statistical analysis. Finally, the interaction volume of the electron beam (and the scattering to generate the X-ray signal) may be substantively different to the volume that generates the bands within the diffraction signal. For the same map point the information within each data type may represent different volumes of matter. To address these challenges, in the present work we explore a weighted PCA approach, prior to VARIMAX rotation, using appropriately prepared and standard deviation normalised EDS and EBSD data.

In this new approach, we have three critical aspects to select in our weighted PCA and VARIMAX rotation method: (1) background correction and data normalisation prior to statistical treatment; (2) the number of RCCs to retain and rotate, corresponding to under- or oversampling of the data, and the variation in the signal for each phase and orientation; (3) the weighting of the signals, dependent on the variance of the EBSD and EDS information, as well as the number of channels in each data set.

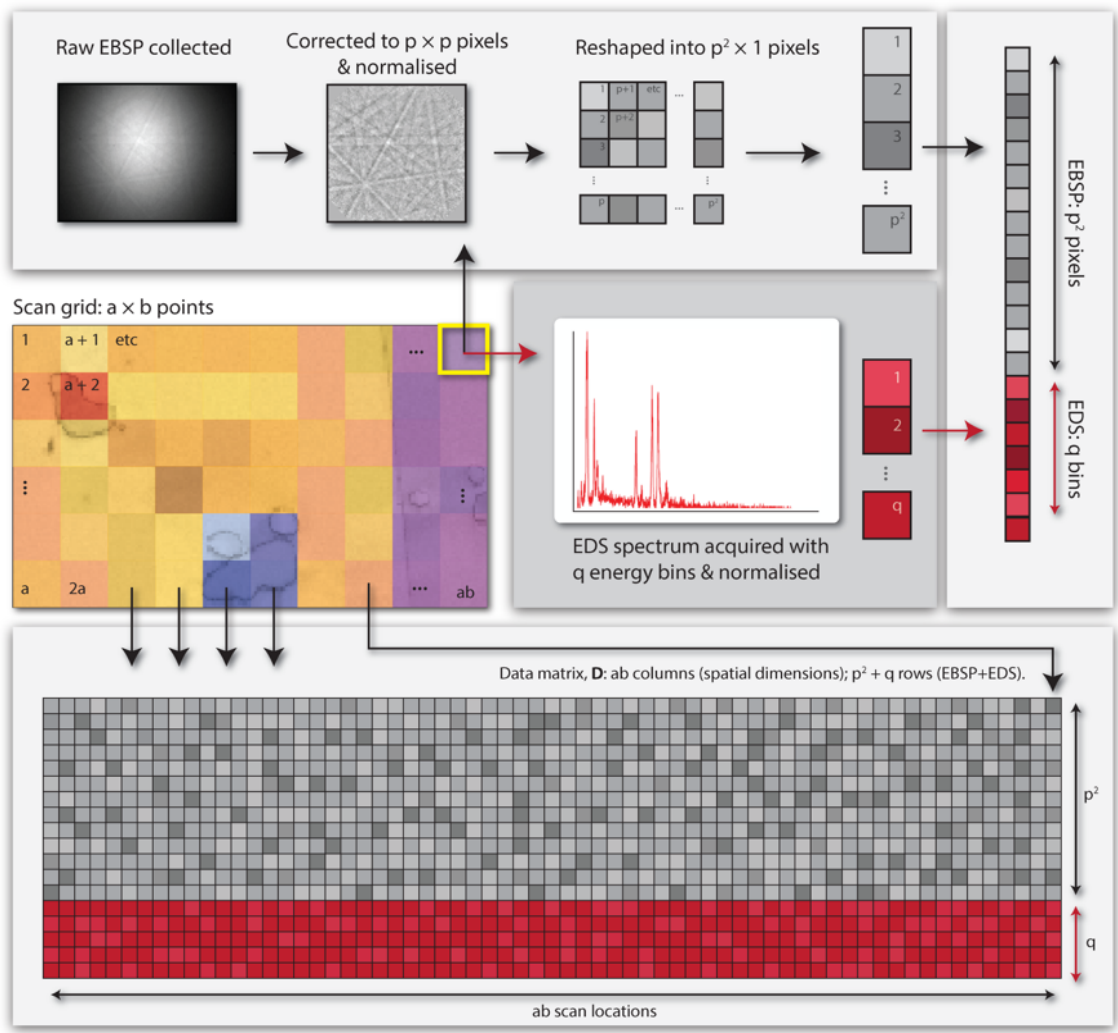

Figure 1: The work flow for construction of the data matrix, **D,** used for the weighted PCA method.

In addition to amplification of signal-to-noise, our statistical approach has computational advantage. Instead of characterising each signal independently, we can use the weighted PCA with VARIMAX rotation to select a reduce set of characteristic signals to quantify. We have a few options; we can directly quantify each characteristic label, but this has to be performed with care as the normalisation and statistical reduction may induce uncertainty (especially where discrete peaks are being quantified in data associated with EDS spectra). We can also use the labelled regions to re-generate amplified means, using averaging, to group together physical data that represents similar phases. Here we explore these two approaches.

Upon classification of scan points to an RC-EBSP and a RC-spectrum, computationally expensive analyses can be applied to a reduced dataset. A ~40,000 point map can be reduced to a few hundred RC-EBSPs and RC-spectra each with a superior signal-to-noise ratio than a single EBSD pattern or EDS spectrum. In this work we analyse the EBSD patterns using the refined template matching approach developed by Foden *et al* [12] with a selection of possible candidate structures. Similar methods have been developed by Ram *et al* [15,16]. Adopting a template matching approach allows us to utilise the fine detail in RC-EBSPs (weakly reflecting bands, band widths, *etc*) that PCA is able to extract. A Hough/Radon-based method would not see significant benefit from this approach, as it is based on comparing angles between the most prominent Kikuchi bands to an interplanar angle lookup table for candidate phases. More structures may be template matched to the reduced set of RC-EBSPs than would be viable for matching to the full experimental dataset, permitting greater confidence in the phase assignment. Characteristic spectra can then be used to quantitatively probe the chemistry of the classifications, and statistically robust comparisons between structure and chemistry can be made. The average and RCC EDS spectra are analysed with commercial EDS analysis software.

## 2. Materials & Methods

### 2.1 Experimental

The alloy characterised in this work is part of a development series of Co/Ni-base superalloys, engineered for high temperature gas turbine applications. It is intraganularly dual-phase, with approximately 55% L1$_2$ gamma-prime volume fraction in a face-centred cubic (FCC) gamma matrix. These two phases have very similar chemistry at the SEM length scale. The EBSD patterns are also extremely similar and can both be indexed with the FCC phase. Refractory element precipitates, *a priori* believed to be carbides, decorate the grain boundaries. A variety of alloying elements are used: Al, Ta and W for gamma-prime stabilisation; Cr for oxidation resistance; Mo for solid solution strength; small additions of C, B, and Zr for grain boundary precipitation. Further details of alloy development have previously been provided elsewhere [24–26].

A Zeiss Gemini Sigma300 FEGSEM equipped with Bruker e⁻Flash$^{HD}$ EBSD detector and XFlash 6160 EDS detector was used for this work. A dataset was captured with 20 kV accelerating voltage at ~ 10 nA and 21.5 mm working distance, with the sample tilted to 70° with respect to the sample being perpendicular to the incident beam. A step size of 100 nm was employed with a pixel time of 8.3 ms. 200-by-150 px EBSD patterns were collected at 16 bit depth, and EDS spectra were captured with 2048 energy channels at 100 eV resolution. The captured data was extracted and stored in a HDF5 file for processing and analysis.

EBSD patterns for each point were processed in MATLAB, with background correction, radial cropping, and hot-pixel/split-chip fixes performed using the AstroEBSD package developed and presented by Britton *et al* [27]. EBSD patterns were originally captured with an aspect ratio of 4:3, but these were cropped to squares to simplify the refined template matching indexing (as discussed by Foden *et al* [12]) prior to creation of the data matrix. EDS spectra were processed in MATLAB. The only pre-processing performed on the spectra was background subtraction and standard deviation normalisation, which we discuss further.

### 2.2 Data treatment and PCA operation

At each map point the corrected EBSD patterns are vectorised and EDS spectra appended, presented in Figure 1. EBSD patterns were background corrected using the AstroEBSD MATLAB package, in which each pattern is divided by a 2D fitted gaussian. Patterns are then centered (mean set to zero and standard deviation set to one). EDS spectra were background corrected using eSprit 2.1 to remove the Bremsstrahlung, then divided by their standard deviation similarly to the EBSD patterns. In the Data Matrix, **D**, each column of data then contains the background corrected and variance normalised EBSD and EDS signals for each measurement point. Each row is the signal for a particular pixel in the binned EBSD pattern or a particular energy in the EDS signal. This Data Matrix follows the formulation of Wilkinson *et al* [13]. The (square cropped) EBSD vector consists of $p^2$ pixels. The EDS signal consists of $q$ bins. **D** therefore contains $p^2 + q$ rows. The Data Matrix contains measurements from *a*-by-*b* points now populating each row, and accordingly the matrix has *ab* columns. Each column is a PCA 'object': a full set of EBSD and EDS measurements ('variables').

The action of PCA and the geometric interpretation of variance (or standard deviation) weighting are presented in Figure 2. A specified number (*n*) of orthogonal principal components are calculated *via* a least-squares singular value decomposition (SVD). These high-dimensionality vectors (in measured variable space) represent the directions in the dataset that explain the most variance. We calculate 'scores' of the components for each object. Scores are the orthogonal projections of the principal components onto the position vector of an object in high-dimensional variable space. They represent how strongly a set of measurements is represented by a principal component. Each object (scan point) has a score for each principal component. In the case of Figure 2b, we scale up variables 2 and 3 to new variables 2a and 3a. This increases the variance in these directions. The dataset is extended in the direction of PC1a, and the object is proportionally linearly loaded more by PC1a than it is by the unscaled PC1.

As previously described, each EBSD pattern and EDS spectrum is normalised with respect to its standard deviation before concatenation and insertion into the data matrix as a column. This is required in order to retain meaningful principal components, and is known to be an important aspect of data pre-treatment due to the variance seeking nature of the PCA parameter fitting process [20].

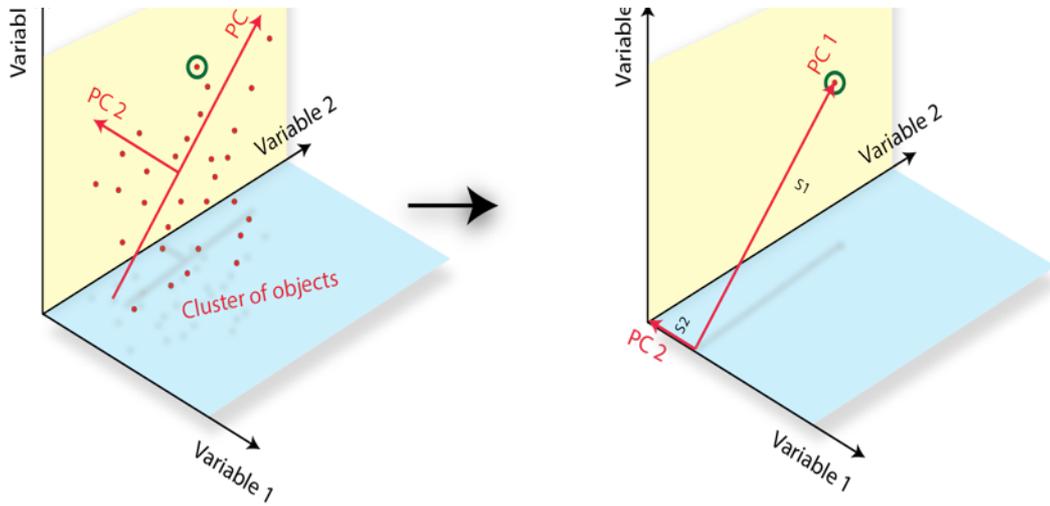

**(b)** Magnified variance of variable 2 and 3 (transformed to 2a, 3a):

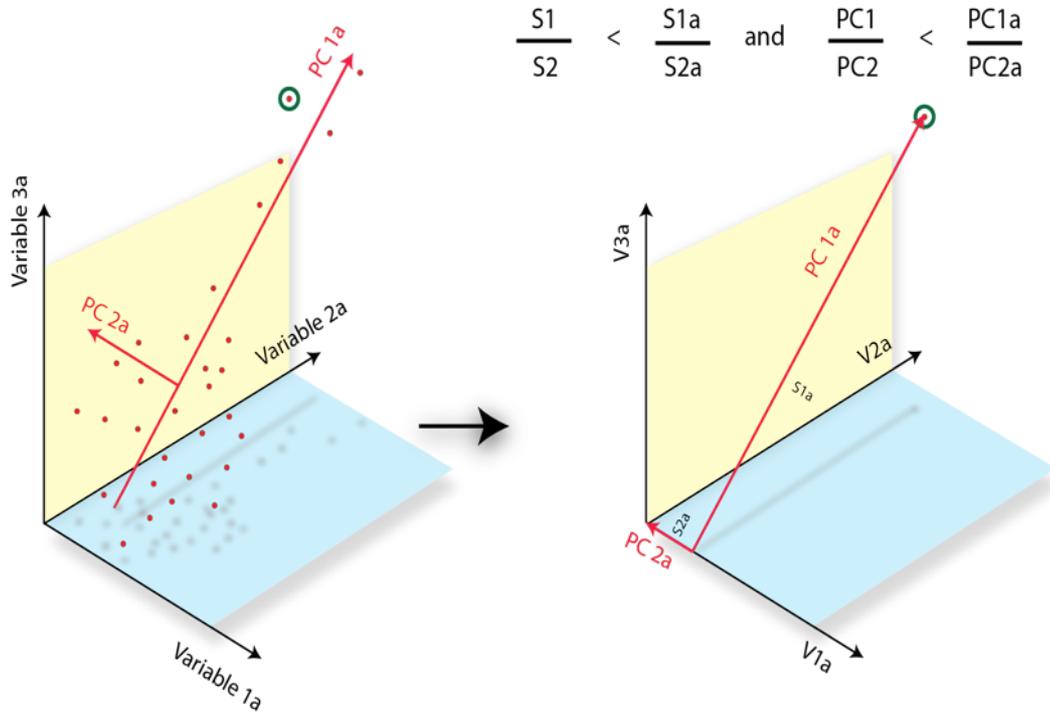

Figure 2: Action of PCA for a schematic dataset with many objects and three variables. (a) shows how the PCA reduces the data set to show strong variation along one principal axis, which may not be an axis of the initial data set; (b) shows how varying the weighting of two combined data sets, which present as information along different axes, can change the variance and therefore the separation of the data sets. Note that this is a simplified schematic for the purposes of visualisation, as our datasets contain tens of thousands of variables.

Variables with high variance across their observations will dominate the principal components, with large corresponding scores for many objects. This is depicted in Figure 2b. As we wish to control the weighting of the principal components with respect to EBSD or EDS, it is essential to normalise the variance of the different variable types upon the data matrix's construction. Without this normalisation of variance for the EBSD pattern and EDS spectrum separately, our weighting parameter would act non-uniformly on each column of **D**, leading to a confused scoring output[1].

---

[1] Note that we do not normalise each row of the data matrix with respect to the variance of the row, as we do not want to treat each channel and pixel number as totally independent measurements. Normalising the variance of each individual

The matrix of principal components, $\mathbf{C}$, and that of their scores at each point, $\mathbf{S}$ are given by:

$$\mathbf{D}_{[p^2+q,ab]} = \mathbf{C}_{[p^2+q,n]} \, \mathbf{S}^T_{[n,ab]}$$

The parameter $n$ is the number of principal components to be fitted, and takes a maximal value of $p^2+q$. $\mathbf{C}$ and $\mathbf{S}^T$ are calculated from the singular value decomposition of $\mathbf{D}$, which is itself the eigendecomposition of the covariance matrix $\mathbf{DD}^T/(r\text{-}1)$, where $r$ is the data matrix's rank. We contain the variance (singular values of the SVD) in $\mathbf{S}$, leaving the principal components themselves with unit length. After Wilkinson *et al* [13] a VARIMAX rotation, $\mathbf{R}$, is then employed, such that:

$$\mathbf{D}_{[p^2+q,ab]} = \mathbf{C}_{[p^2+q,n]} \, \mathbf{R}_{[n,n]} \, \mathbf{R}^T_{[n,n]} \mathbf{S}^T_{[n,ab]}$$

The matrices $\mathbf{C}$, $\mathbf{S}$ and $\mathbf{R}$ are numerically calculable in MATLAB using the statistics and machine learning toolbox. The VARIMAX rotated characteristic components (RCCs) are held in the rows of the matrix $\mathbf{CR}$, with the corresponding scores for every point given by the rows of $\mathbf{SR}$. RC-EBSPs and RC-spectra may then be re-constructed from the first $p^2$ and final $q$ rows of $\mathbf{CR}$ respectively. An $a$-by-$b$ assignment map can be constructed with the same spatial dimensions as the original scan grid. Each point is assigned a number, $m$, corresponding to the RCC with the highest score. For each scan point's corresponding row in $\mathbf{SR}$, $m$ is the number of the column that takes the greatest value. Each point is thus classified to one of $n$ labels to construct an assignment map. Each label is associated with a characteristic EBSP (RC-EBSP) and spectrum (RC-spectrum).

As we have constructed this algorithm, the weighting term we introduce acts to reduce the variance of the EBSD variables, magnifying the relative standard deviation observed for each EDS variable (across all points in the map, columns of the data matrix) by a factor $w$. This has the effect of increasing the influence of the EDS variables on the principal components through the action of Figure 2b. In practice, this means that there is an observable transition from EBSD-dominated through to EDS-dominated behaviour as $w$ is decreased, this is presented and discussed in sections 3.2 and 3.3.

Upon construction of the data matrix $\mathbf{D}$, the PCA algorithm we employ has two parameters that require selection. These are: the number of components that we choose to retain for the VARIMAX rotation, $n$, and the EBSD standard deviation weighting, $w$. Selection of these will be discussed subsequently. To make the processing tractable on a reasonable computer, there is a requirement to divide the full region of interest into smaller tiles so that there is sufficient memory available for the SVD algorithm (we need to hold in memory a rank $r$ square matrix). 3-by-3 tiling was employed for the datasets presented in this work. The RAM requirement for processing each tile of this dataset is 58 gigabytes. Principal components and RCCs are calculated for each tile fully independently, with slight consequence discussed in section 3.3.

## 2.3 Analysis of PCA output

Reshaping the first $p^2$ rows in all $n$ columns of $\mathbf{CR}$ into $p$-by-$p$ images reconstructs $n$ RC-EBSPs. The final $q$ rows for all $n$ columns correspond to RC-spectra. These are separately analysed and quantified. The reduced dataset of characteristic patterns and spectra has a superior signal-to-noise ratio to the experimental measurements. We can analyse the data in the form recovered from the weighted PCA and VARIMAX rotation, but we may have issues for instance where two carbides have the same chemistry and phase (i.e. similar EDS spectra) but different crystal orientations (varying EBSD data). Therefore it is useful for us to analyse the labels in more detail.

We label our data first using EBSD pattern analysis. In this work we apply the Refined Template Indexing approach developed by Foden *et al* [12] to assign phase and orientation to each of the point labels. This method involves cross-correlation of test EBSPs (in this case the RC-EBSPs) with a database of library patterns, sampled with the fundamental zone of SO(3) space for each crystal structure with a specified angular frequency. The master patterns were dynamically simulated using Bruker DynamicS [11,17] and reprojected in MATLAB using the pattern centre calibrated from the Ni-rich matrix. Sampling of SO(3) was performed with a frequency of 7° and refinement

---

measurement across all scan points would reduce or completely eliminate the prominence of features such as EDS peaks and Kikuchi bands in the principal components.

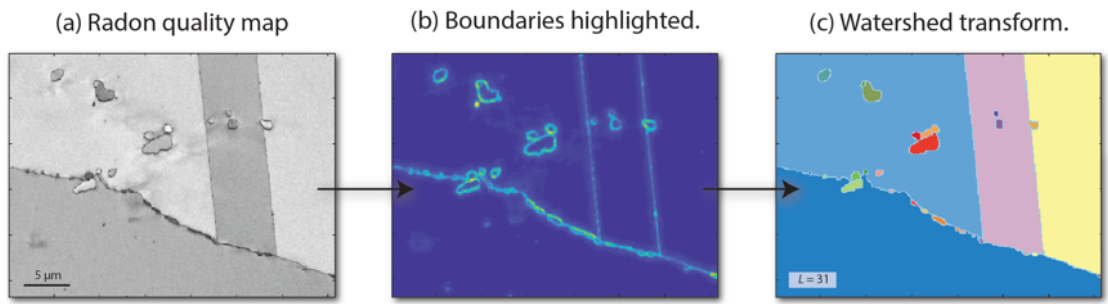

Figure 3: Filtering and watershed transform of Radon quality map to determine a value of *L* for VARIMAX rotation.

was used to upsample the orientations. Templates were generated using five input candidate crystal structures were considered for RTI template matching: FCC Co, $M_{23}C_6$, $M_2C$, $M_6C$ and MC selected from the literature [5,28–30]. CIF files and unit cell visualisations are included in the supplementary information. The RC-EBSPs in these datasets were indexed as FCC Co, $M_6C$ or MC (selecting each phase based upon the highest ranked cross correlation value and scrutiny of pattern matching).

For the EDS data, we export the data in a format that can be analysed directly in Bruker eSPRIT 2.1, and quantify RC-spectra using a ZAF correction algorithm that accommodates the 70° sample tilt required for EBSD. We additionally compute and compare the average measured EDS spectra from all points assigned to a given RCC label.

## 3. Optimising weighting and VARIMAX rotation

Here the effect of two important input parameters, *n* and *w,* will each be discussed. These correspond to the number of principal components we retain and the weighting factor for EBSD information that leverages PCA in favour of either EBSD of EDS.

### 3.1 Retention of components

While we can select the number of principal components worth retaining, *n*, by hand based upon a qualitative assessment of the appearance of our final micrographs, it is also useful to explore whether there are quantitative, or semi-quantitative assessment processes that can guide our selection. This is a problem that has been considered extensively in the data science literature [20,31–34].

Here, we perform a VARIMAX rotation on the first *n* principal components and expect a single label to represent structural and chemical information contributed from one grain. If *n* is greater than the number of grains in the AOI, the additional labels will correspond to sub-grains (which may be advantageous but will reduce the signal to noise ratio of the RCCs). If *n* is too large, the calculated scores of nearby (and similar) points will be high for multiple labels and there will be noise in the assignment map. If *n* is too low, the grains will not be properly segmented and information will be lost as we have over-reduced the dataset. This work presents two approaches for selecting *n*. The first involves counting the number of grains in the measured EBSD-based Radon quality map. The second imposes a limit on the contribution of the first *n* principal components to the total variance of the dataset.

#### 3.1.1 Counting grains to select *n*.

A reasonable value of *n* is an estimate of the number of grains in the EBSD-based Radon quality map, *L*. This was calculated with Bruker eSPRIT 2.1 and is an essentially 'free' microstructural image that is spatially consistent with the EBSD and EDS measurements. Several approaches have been reported for counting the number of grains in a microstructural image. These include the application of an 'H-concave' transformation to channelling contrast forescatter electron images (with subsequent refinement) developed by Tong *et al* [35]. This grain counting step is employed to select the locations of a dramatically reduced number of EBSD patterns for an orientation map. The 24-bit information depth of the RGB colour image constructed from the forescatter diode intensities allows segmentation of scan points into labels of similar colour and contrast. A subsequent refinement step where each point is compared to the labels of its neighbours leads to a very accurate image reconstruction. Campbell *et al* [36] utilise a 'Watershed' transform to identify and distinguish phase fractions and morphology in grayscale SEM images of Ti-6Al-4V. This algorithm treats an image as a topographic region of intensity basins. A labelled source is placed at each local minimum and allowed to flood the image. Image segmentations are delineated where floods from different sources meet [36–38]. When applied to the local gradient of a microstructural image (in order to highlight boundaries and leave grain interiors with low intensity), reasonably accurate intensity classification can be achieved. However, the algorithm has a tendency to over-segregate and assign too many labels.

For its speed and simplicity the Watershed algorithm was selected to quickly identify a value of *L*. The image processing steps are presented in Figure 3a. Starting from the EBSD-based Radon quality map (Figure 3a) a series of local averaging filters, ending with the image complement of a standard deviation filtered image and local minima flattening, are used to highlight boundaries (3b) A watershed transform then assigns labels to different regions (3c) following the topographical method of Meyer [37]. We then select *n* equal to *L* principal components, and then we perform the VARIMAX rotation.

This approach is fast and provides a reasonable estimation of *n* for an area of interest. As will be shown subsequently the watershed algorithm significantly oversamples the subsets, especially where coherent intergranular precipitates with similar orientation and chemistry are counted separately by the watershed algorithm. This leads to too great a value of *n* being selected and the signal-to-noise ratio of the characteristic patterns and spectra are not optimised.

Care should be taken, as the ability for the watershed algorithm to determine the number of components will depend on the types of features presented. In this example, we are exploring an annealed Co/Ni matrix and so there is minimal contrast in that region, but we want to focus on the number of signals from the carbides. In noisier Radon quality maps it seems likely that the watershed algorithm would perform significantly worse. This could potentially be somewhat mitigated for a wider variety of datasets by further *ad hoc* convolutions and image pre-processing.

### 3.1.2 Limiting variance contribution of the $n^{th}$ principal component.

Conventional PCA relies on the relatively subjective identification of an inflection in a Scree plot (for an example, see Figure 4) to estimate the number of components that significantly and sufficiently describe the variance of the dataset [20,33]. The Scree plot describes the explained variance contribution to the dataset as a function of principal component, derived from the corresponding eigenvalues of the covariance matrix. In this plot the principal components are ordered by their contribution to the variance, with the first being the highest contributor. The inflection point in a Scree plot may indicate where the principal components cease to add new substantive value to a description of the data, but often the inflection is unclear.

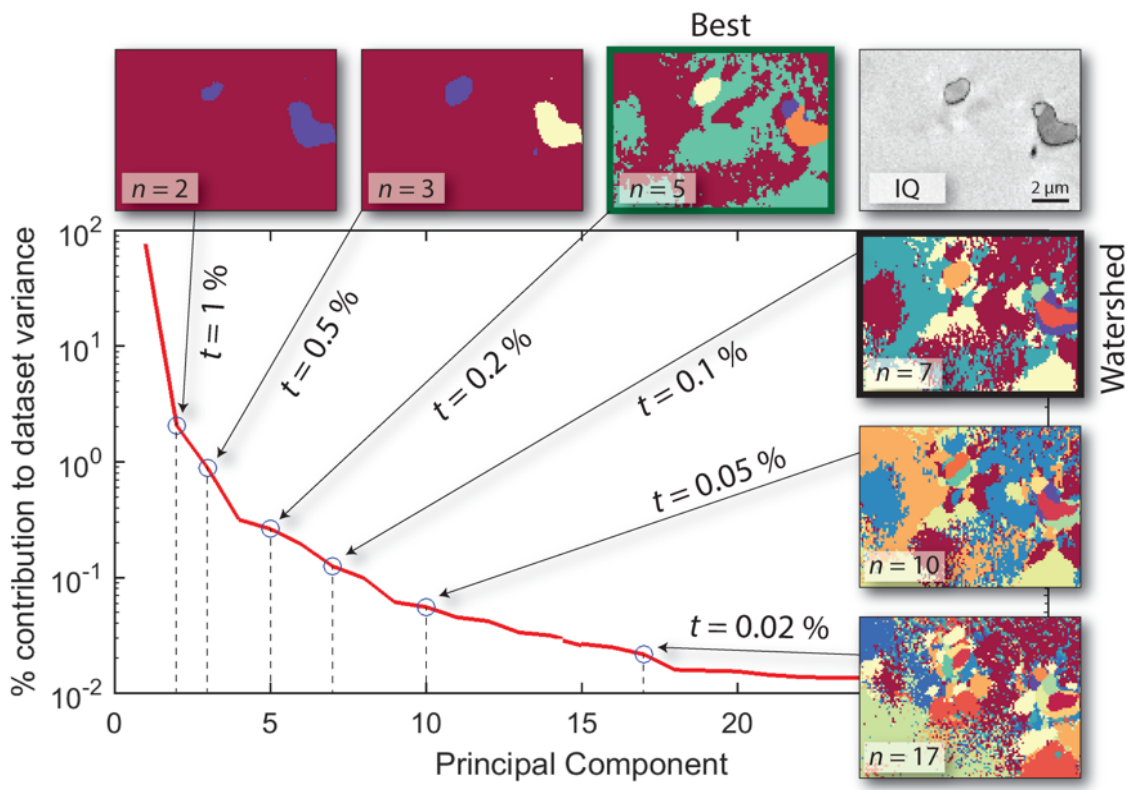

Figure 4: Selection of *n* based on applying a cut-off tolerance, *t*, for percentage contribution to total dataset variance of the $n^{th}$ principal component. The third PC contributes < 1% of the total dataset variance. The 18$^{th}$ PC contributes < 0.02%. The watershed algorithm discussed in section 3.1 selected seven components for this AOI.

Alternatively, the total explained variance for a certain number of retained principal components itself provides a metric for the utility of the retained components. This idea can aid in selection of $n$, the number of principal components to retain for VARIMAX rotation and generation of characteristic patterns (RC-EBSPs) and spectra (RC-spectra). We can set $n$ as the integer for which explained variance of the $(n+1)^{th}$ principal component falls below a certain threshold, $t$, for example 0.2%. All $n$ principal components then contribute a variance proportion greater than $t$ to the dataset. Information contained in the remaining principal components is discarded (i.e. this is considered as noise). The short circuit caveat around this selection rule is that $n$ must be greater than or equal to 2 for VARIMAX rotation, so in some cases (with high initial $t$) we are forced to select $n$ with variance contributions greater than $t$. We can therefore choose $n$:

$$VarianceContribution\{ (n+1)^{th} PC \} < t$$

$$If\ n < 2, set\ n = 2.$$

This is a relatively crude selection algorithm, but we demonstrate that it works. More sophisticated methods discussed by Raîche et al [33] using non-graphical metrics such as the Scree test optimal co-ordinate or Scree test acceleration factor could be employed. These values of $n$ correspond to analyses of the gradient of the explained variance (equivalently the eigenvalues of the covariance matrix) as a function of $n$. We did not find these latter criteria suitable for our datasets, and they would systematically select the maximum statistically allowable number of components as described by Kaiser's rule, which stipulates that the eigenvalue of the $n^{th}$ retained component must not be less than one [39].

Considering the first tile of the full AOI, Figure 4 shows the Scree plot for a PCA with EBSD weighting parameter (discussed further in section 3.1), $w$, equal to one, and the resulting assignment maps after VARIMAX rotation for values of $n$ selected with varying variance tolerance criteria, as well as the Watershed algorithm. The variance tolerance limit, $t$, is selected to vary between 1% (leaving two principal components) and 0.02% (leaving 17 principal components).

The quality of the label assignment to each point can be quantified by normalised cross-correlation of measured EBSD patterns and EDS spectra with a point's corresponding characteristic RC-EBSP and RC-spectrum. Maps of these correlation values are helpful to visualise how this varies across the AOI, and how well different precipitates and grains match to their corresponding RC-EBSPs and RC-spectra. Maps of these correlation qualities (normalised correlogram peak heights at zero lag, $\chi_{EDS}$ and $\chi_{EBSD}$) are presented in Figure 5a. In the presented case for $w = 1$ there is little variation in $\chi_{EDS}$ as dataset variance is dominated by EBSD information. We also consider a quadrature combination of EBSD and EDS correlogram peak heights, $\chi_{comb}$:

$$\chi_{comb} = \sqrt{\chi_{EBSD}^2 + \chi_{EDS}^2}$$

Four possible metrics for the classification quality of an analysis are suggested. These are each plotted as a function of the variance of the final, $n^{th}$, component in Figure 5.

- <u>Metric 1</u>: The proportion of points that satisfy $\chi > 0.95\ \chi_{max}$ (for $\chi_{EBSD}$ and $\chi_{EDS}$) - to be maximised, Figure 5b.
- <u>Metric 2</u>: The proportion of points that satisfy $\chi_{EBSD} < 0.7\ \chi_{EBSD,max}$ or $\chi_{EDS} < 0.9\ \chi_{EDS,max}$ - to be minimised, Figure 5c.
- <u>Metric 3</u>: Mean value of $\chi_{EBSD}$, $\chi_{EDS}$ or $\chi_{comb}$ − to be maximised, Figure 5d.
- <u>Metric 4</u>: Standard deviation in $\chi_{EBSD}$, $\chi_{EDS}$ or $\chi_{comb}$ − to be minimised, Figure 5e.

Selecting the 'best' value of $n$ for an AOI can be made less subjective by choosing a variance tolerance limit that maximises or minimises one of these metrics. When the tolerance limit is relaxed (low $t$, high $n$, with $n^{th}$ component contributing only a small amount of variance) the AOI is oversampled with principal components. This leads to a noisy assignment map (observed for the assignment map in Figure 5a with $n = 17$), as nearby (and similar) points have similar scores for several RCCs. As the tolerance limit is tightened (moving to the right in the graphs of Figure 5b-e), the assignment initially improves across most of the metrics. Percentage of correlogram peak heights close to the maximal values increases (Figure 5b), and percentage much less than the maximal decreases (Figure 5c). The mean increases (Figure 5d) and standard deviation falls (Figure 5e). This initial improvement is attributed to improving the signal to noise ratio of the RCCs, as we reject superfluous principal components. As the tolerance limit is tightened further and $n$ is reduced, the assignment moves past an optimal position. Beyond this point insufficient principal components are included in the VARIMAX rotation. This leads to insufficient (and inaccurate) RC-EBSPs and RC-spectra. Accordingly the cross-correlation peak heights for this dataset fall – the mean decreases and standard-deviation rises. We also observe that the Watershed algorithm oversamples the dataset.

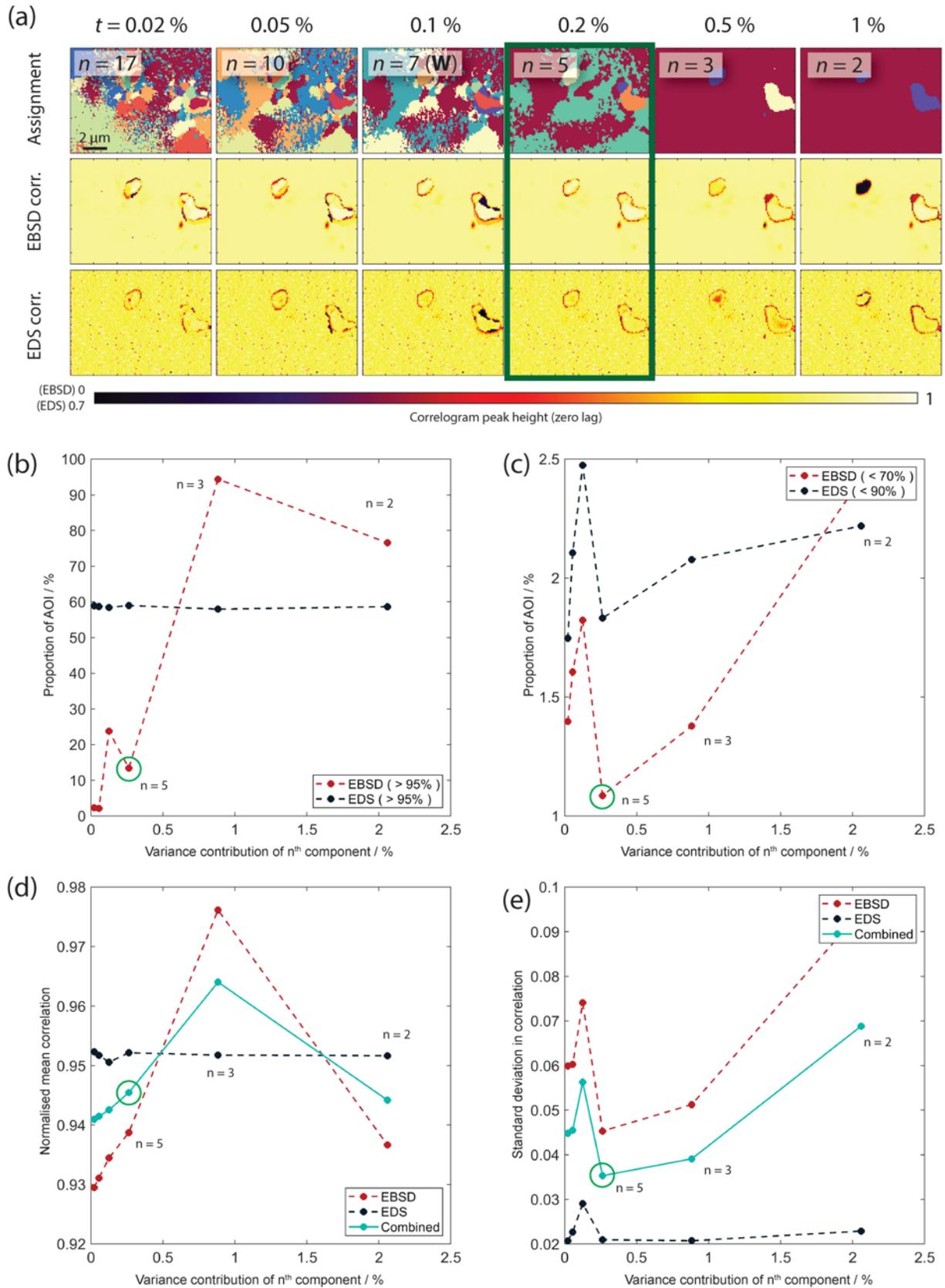

Figure 5: (a) Effect of varying variance rejection percentage on measured EBSD pattern and spectrum cross-correlation peak height with corresponding RC-EBSPs and RC-spectra. Colour is cross-correlation peak height. (b) and (c) show percentage of these EBSD and EDS cross-correlation peak heights greater or less than cut-off proportions of maximal correlation in that analysis. (d) and (e) show mean and standard deviation of the cross-correlation peak heights for EBSD, EDS and the quadrature combination of the two.

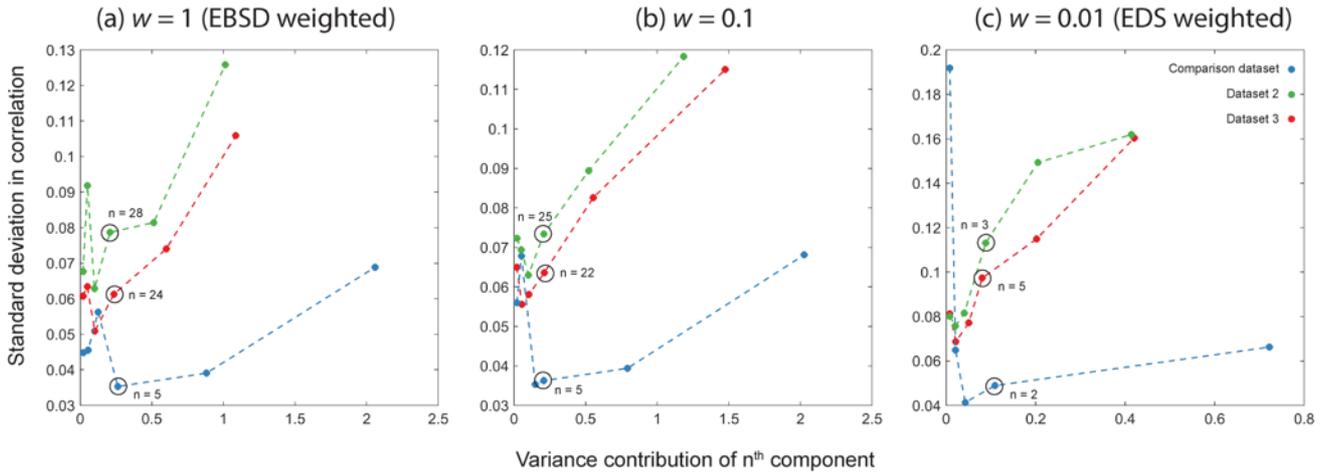

Figure 6: Standard deviation of normalised cross-correlation between measured EBSPs and spectra with their associated characteristic EBSPs and spectra, combined in quadrature. Shown for three alloys and three EBSD weighting parameters. As more principal components are retained, correlation initially improves as AOIs are better matched. After reaching the optimal level correlation degrades as principal components corresponding to noise are included in the VARIMAX rotation. The trend is stable (the minimum occurs in approximately the same place) across independent datasets and weighting parameters. $n$ corresponding to $t = 0.2\%$ is circled in each case.

For the analysis presented in Figure 5 we determine that a choice of $t = 0.2\%$ provides a good assignment. This was selected as the minimum in the standard deviation based metric 4. This $t$ also performs best in metric 2 (Figure 5c) and second best in metric 3 (Figure 5d). Metric 4 appears to be a good choice for deciding the optimal value of $t$ (and therefore selecting $n$ in independent datasets). A choice of $t = 0.5\%$ ($n = 3$) leads to superior metrics 1 and 3 than $t = 0.2\%$ ($n = 5$). However, the assignment map in Figure 5a shows better correlation (both EBSD and EDS) for $t = 0.2\%$ ($n = 5$). Evidently the matrix regions correlate slightly better for $t = 0.5\%$ than for $t = 0.2\%$, raising the mean peak height despite an observed poorer correlation for the precipitate regions. An approach of minimising poor correlation, by either of metrics 2 and 4, is less sensitive to this effect, and $t = 0.2\%$ ($n = 5$) exhibits obvious minima. Furthermore, the percentage difference between measurements of mean $\chi_{comb}$ is far smaller than that for standard deviation of $\chi_{comb}$, providing a more justifiable minimum.

It was found (and is shown in the supplementary data) that metrics 1 and 2 are sensitive to the choice of proximity parameters (here 95% maximal for

EBSD and EDS, 70% minimal for EBSD and 90% minimal for EDS). In contrast, we observe that trends in standard deviation of $\chi_{comb}$ as a function of $n^{th}$ component variance are stable between datasets, choice of $w$, and the specific values of $n$ that the tolerance limits correspond to, shown in Figure 6. Based on the stability of the standard deviations in $\chi_{comb}$, a variance tolerance $t = 0.2\%$ was selected for selection of $n$ in subsequent analysis of the effect of the EBSD weighting parameter $w$.

### 3.2 Leveraging relative EDS variance

The dataset variance contribution of the EDS energy bins, the final $q$ rows of $\mathbf{D}$, is altered *via* a scaling factor in order to bias the PCA in favour of EBSD or EDS information. Without any weighting, the far greater number pixels in an EBSD pattern compared to energy bins in an EDS spectrum (65,536 for a 256-by-256 pixel EBSD pattern, and 2048 channels for our energy-binned spectra) dominate the variance of the dataset unless the former is dramatically scaled down. This is achieved by multiplying the EBSD and EDS rows of $\mathbf{D}$ by $w$ and one respectively [21]. The variance normalisation step employed during the construction of $\mathbf{D}$ separately reduces the standard deviation of the input EBSPs and EDS spectra to one. The weighting multiplication therefore reduces the standard deviations of the input EBSPs to $w$, while that of the input spectra remains equal to one.

The tile AOI presented in Figure 5 was processed with weighting parameter, $w$, varied between 0.001

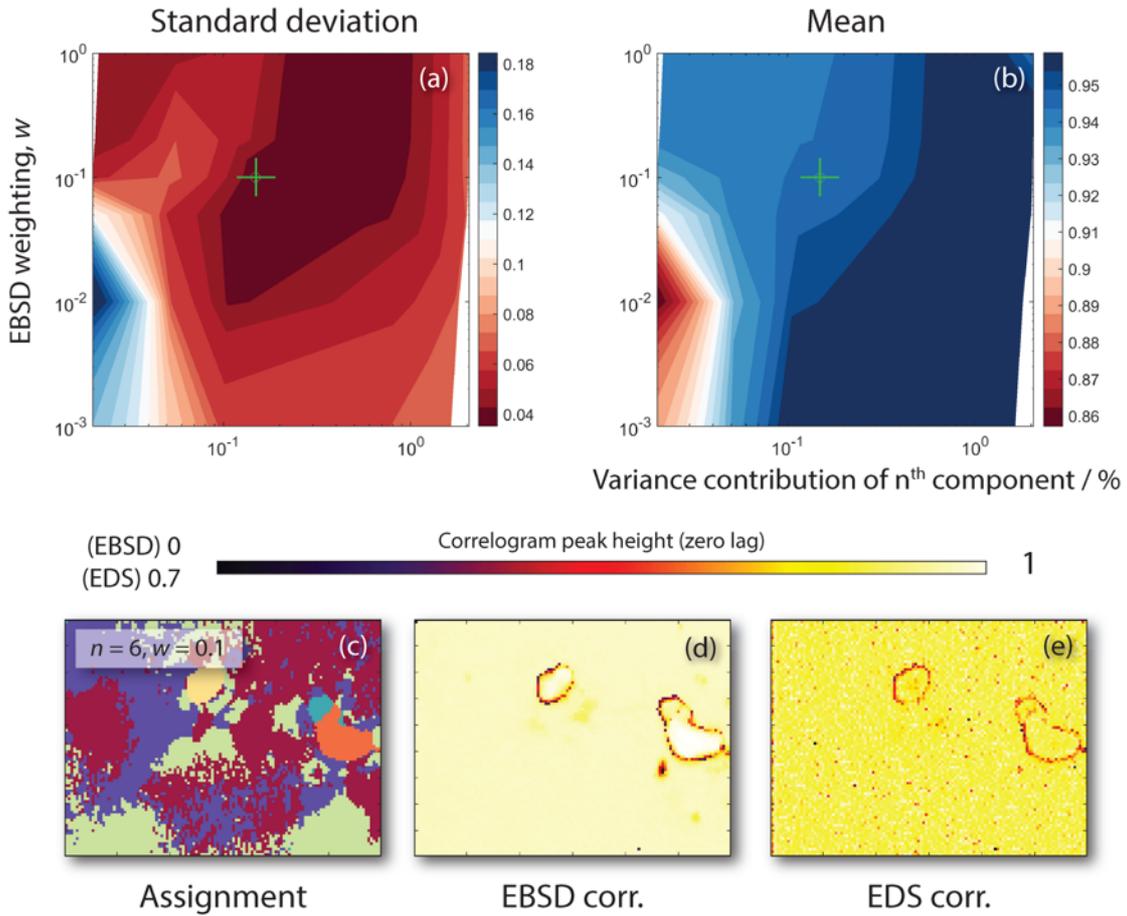

Figure 7: Standard deviation (a) and mean (b) of combined EBSD and EDS RCC/measurement cross-correlation peak heights, $\chi_{comb}$, defined in section 3.1.2. The 'best' assignment map is presented (c), along with correlation peak height maps for EBSD, $\chi_{EBSD}$ (d) and EDS, $\chi_{EDS}$ (e). 36 data points are included in the $t$ / $w$ parameter space maps.

and 1. A variance tolerance limit, $t$, was used to select the number of components to retain, $n$, varied between 0.02% and 1%. Maps of standard deviation and mean of $\chi_{comb}$ as a function of $t$ and $w$ are presented in Figure 7. A local minimum in standard deviation is identified. The corresponding assignment map is included (Figure 7c), along with maps of $\chi_{EBSD}$ (Figure 7d) and $\chi_{EDS}$ (Figure 7e).

The same trends in variance tolerance limit are observed as in Figures 5 and 6, which present $\chi_{comb}$ as a function of the variance contribution of the $n^{th}$ component for a single $w$. As the tolerance limit is tightened and the $n^{th}$ component has to contribute more variance (left to right in 7a, 7b), the standard deviation in $\chi_{comb}$ falls and the mean rises. As in Figure 5, a local minimum in standard deviation is observed as $t$ is varied. Beyond this limit insufficient principal components are retained. As $w$ is increased (tending towards EBSD weighting, upwards in 7a, 7b), standard deviation in $\chi_{comb}$ generally decreases. The mean value of $\chi_{comb}$ increases as $w$ increases for loose variance tolerance limits (AOI oversampling), but at higher $t$ (smaller $n$) there appears to be less of a correlation between mean $\chi_{comb}$ and $w$. Considering the standard deviation in $\chi_{comb}$ (metric 4 of section 3.1.2) as a measure of assignment quality identifies a seemingly optimal combination of $w$ and $t$. Associated label assignment, $\chi_{EBSD}$ and $\chi_{EDS}$ maps are presented (7c-e).

### 3.3 Full dataset processing and assignment artefacts

The PCA approach achieves the desired signal-to-noise improvement for poorly diffracting particles such as $M_6C$ carbides. Example measured, label (RCC), and matched simulation patterns for the pseudo-FCC matrix and $M_6C$ carbide are presented in Figure 8.

Considering the full AOI of nine tiles, Figure 9 presents the dataset processed with $w = 1$ (EBSD weighting) and $w = 0.1$ (EDS weighting, identified as the local minimum in standard deviation of $\chi_{comb}$ in section 3.2). RC-EBSPs were indexed using the Refined Template Matching procedure [12], and RC-spectra were quantified with Bruker eSprit 2.1.

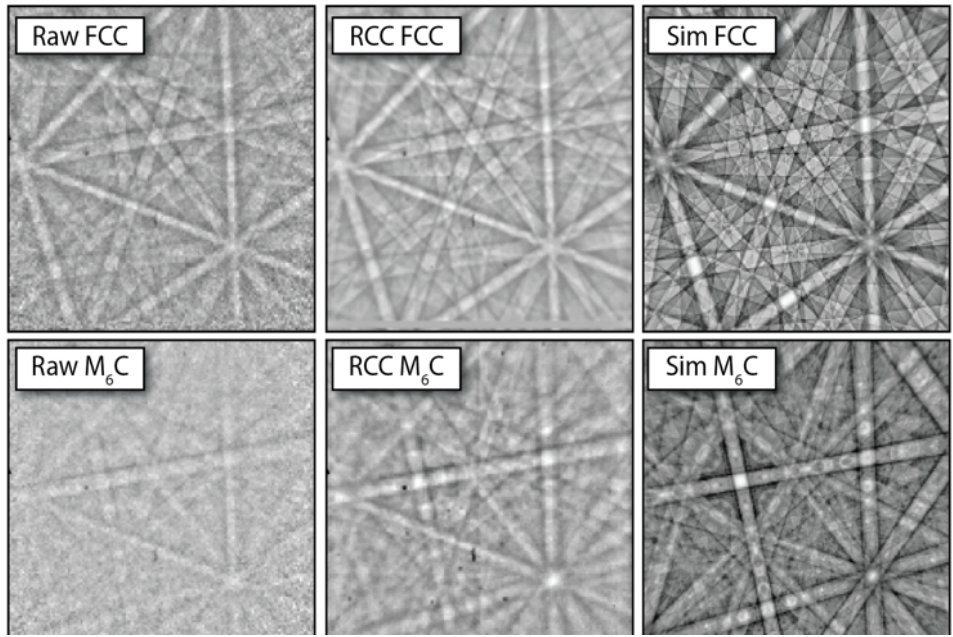

Figure 8: Comparisons of example raw, rotated characteristic component (RCC), and template matched dynamically-simulated EBSPs for the pseudo-FCC Ni/Co matrix and the $M_6C$ carbide phase. This demonstrates how the method amplifies the quality of the minor phase substantially, which assists in unambiguous classification.

When the EDS weighting is high (Figure 9b,d,f,h), label assignment is dominated by the magnified variance of the EDS spectrum energy bins. To qualify this, we explore the interaction volume using Monte Carlo simulations and the continuous slowing down approximation, this provides an indication of the interaction volume of the electron beam and X-ray generation. We note that the generation of the background signal for EBSD is likely to be smaller than predicted from the CSDA-approximation as the energy of the electrons that form the Kikuchi bands is constrained to be closer to the primary beam than the CSDA predicts [40]. The interaction volume of backscattered electrons in this system, simulated with CASINO at 70˚ sample tilt and detailed further in the supplementary information, is at most 100 nm. That for secondary (X-ray generating) electrons is significantly larger. The Monte-Carlo simulation suggests that secondary electrons from the FCC matrix are generated up to perpendicular depths of 1 μm. Carbides exhibit even larger volumes, with $Cr_6C$ and ZrC carbides interacting up to 1.4 μm and 1.7 μm respectively.

The consequence of large EDS interaction volumes is that at high magnification (where scan step is much less than the coarser technique resolution) two adjacent scan positions with measurably different crystal structure may exhibit very similar EDS spectra. The position vectors of these observation sets in variable space are very similar, despite differences in the measured EBSD pattern, due to the demagnification of EBSD pattern variance in this analysis. Effectively this leads to a loss of spatial resolution in label assignment, and as highlighted at position A, the possibility of missing fine precipitates from the classification. It can be seen in the EDS weighted assignment map Figure 9d, that an elongated characteristic region of C enrichment and different chemistry, follows the grain boundary. The greater number of matrix points in this region dominate the principal component-EBSP, and the RC-EBSP for this region indexes as FCC Co.

Another artefact we observe in the assignment is highlighted at position B. In the EDS-weighted PCA we observe the upper region of a precipitate (MC carbide) grain is assigned a different orientation to the remainder below. This artefact is the confluence of two method limitations. The need to tile the dataset due to the significant memory requirement of the SVD algorithm means the upper and lower regions have to be assigned labels independently. This is not an issue for matrix regions, as there is a sufficient population (and therefore dataset variance contribution) to assign noise free and appropriate RC-EBSPs and RC-spectra. However, when the PCA is EDS-weighted there is insufficient EBSD variance (due to a small population of points in the upper half of the mis-assigned precipitate) to provide a label that contains sufficient orientation data for this region to be separated from the nearby second grain of chemically similar MC carbide. This results in the upper region of the split precipitate being assigned the correct chemistry and phase, but an incorrect orientation. This could be compensated for by relaxing the variance tolerance limit imposed on this analysis or an alternative sampling strategy to retile for small segments towards the tile edge.

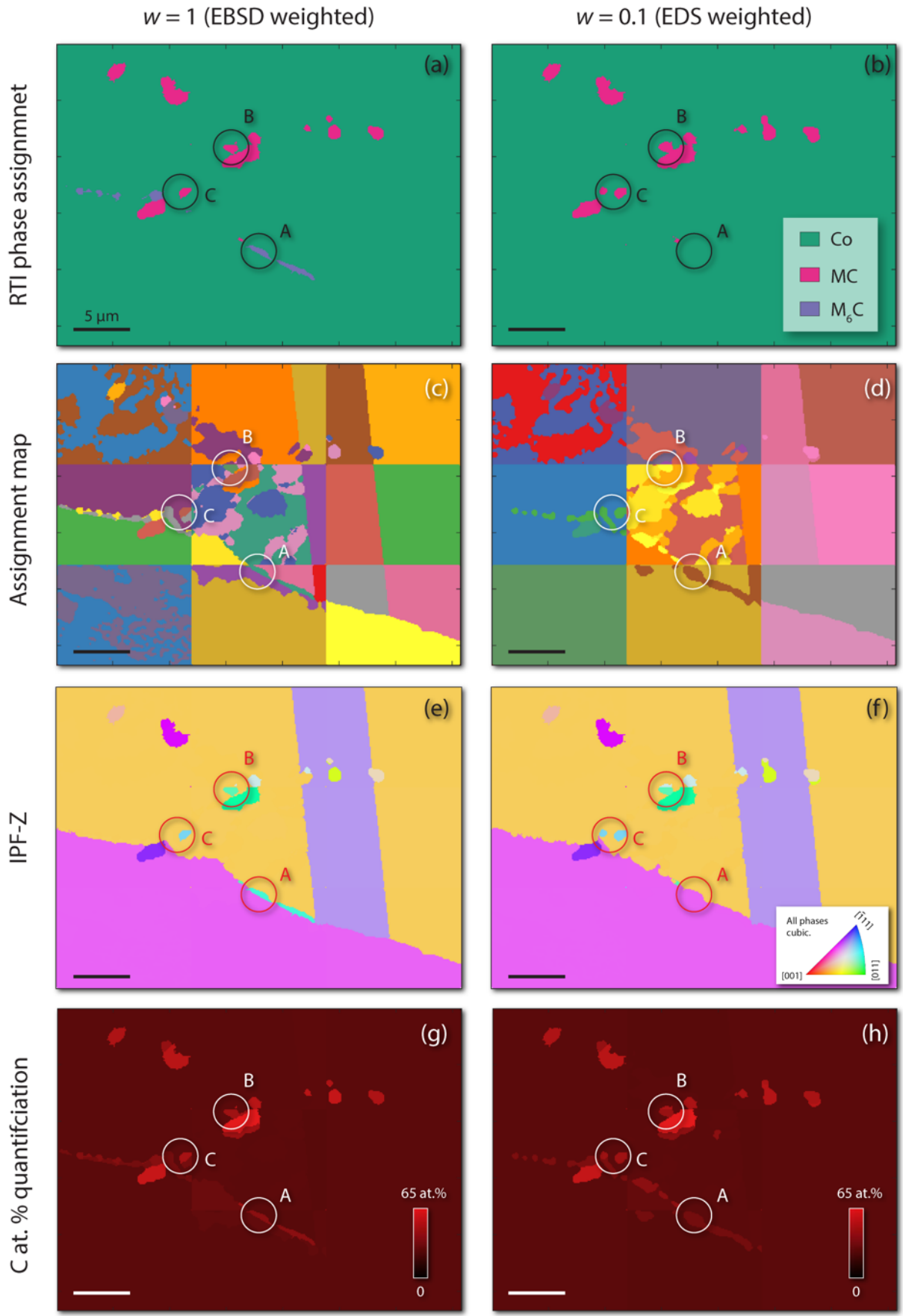

Figure 9: Comparison between phase assignment (a-b), label assignment, with arbitrary colouring (c-d), IPF-Z – out of plane (e-f), and C at.% from the RC-spectra quantified with Bruker eSprit 2.1 (g-h), after processing with $w$ = 1 (EBSD weighted) and $w$ = 0.1 (EDS weighted). Both analyses were performed with a variance tolerance limit $t$ = 0.2 %.

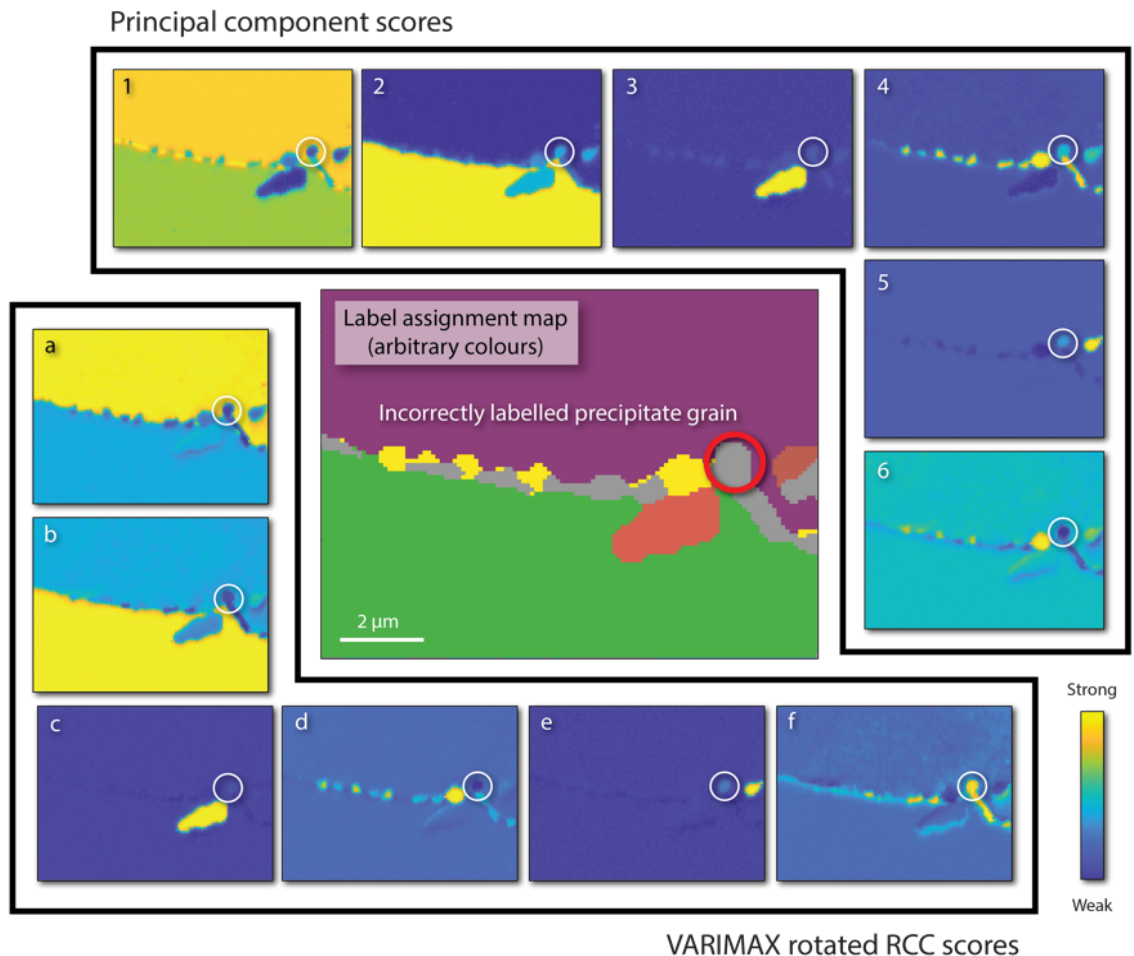

Figure 10: Second tile AOI (with arbitrary label colouring) of Figure 9 with EBSD weighting, *w* = 1. The mis-labelled precipitate grain (point C in Figure 9) is highlighted. PC scores and VARIMAX rotated RCC scores are presented. Principal components are ordered by their contribution to dataset variance, but RCCs are calculated to contribute equal variance. Score colour map is unnormalised between maps to show contrast[1]. No principal component strongly contributes to the highlighted grain (1-6), and the VARIMAX rotated component with the highest score (f) includes significant signal from several other precipitates and matrix. This results in the corresponding RC-EBSP being dominated by FCC Co signal, and indexing accordingly (point C in Figure 7).

A third artefact is highlighted at position C. In the EBSD weighted Figure 9f, one MC carbide precipitate is identified and labelled. Two are assigned in EDS weighted Figure 9e. In this case the applied variance tolerance limit of 0.2% is insufficient to separate this region from the surrounding matrix. The region is highlighted in Figure 10, and score maps for the first six calculated principal components and the VARIMAX rotated

RCCs are presented. The precipitate is mis-indexed in this analysis due to insufficient sampling of principal components. No PC strongly contributes to the precipitate grain (Figure 10, highlighted in 1-6), and little signal from this region is included in the VARIMAX rotation and calculation of the six RCCs. This results in the precipitate being labelled with an FCC Co matrix dominated RC-EBSP, and eventually indexed as such.

## 4. Discussion

Easy access to advanced statistical treatments enable us to treat microscopy data as a 'big data' problem and we are likely to see increased use of these approaches. We have illustrated that consideration of the data modality (e.g. physical processes to generate the signal, combined with the statistical variance of each data type) provides improved confidence in their use to reasonably, and usefully, segment large data sets.

The present work presents a limited size of region with only a few domains, but testing using a number of other (lower magnification) maps indicates that our variance tolerance limit is a good indicator of the number of domains - grains, sub-grains, and precipitates - within Ni-based maps containing many more of these features. We hope to explore this in future work, where we will use the method to explore substantially larger combined EBSD and EDS maps.

Our combination of EDS and EBSD signals together using a weighted PCA approach, with subsequent label identification and characterisation, improves phase characterisation within the scanning electron microscope. To combine these modalities, we have to select an appropriate data processing pipeline to provide robust data mixing, with subsequent selection of an appropriate number of components for retention prior to VARIMAX rotation. This is required to inform correct identification of the labelled regions. We achieve this through selection of suitable values for the two independent hyperparameters $w$ and $n$ (the latter varied through $t$, the variance tolerance limit). Here we review our approach and discuss applications and potential utility of the technique.

## 4.1 Parameter choice and data-type leverage

We have shown that a PCA algorithm may be biased towards obtaining RCCs through identifying the strongest signals in either EBSD or EDS information. An EBSD-weighted PCA exhibits a finer effective spatial resolution in label assignment, due to the smaller interaction volume for electron backscattering than for X-ray generation. From this we obtain RC-EBSPs and RC-spectra identified from structurally contrasting points in an AOI. We can also leverage PCA in favour of EDS spectrum dissimilarity and identify RC-EBSPs and RC-spectra accordingly. In this case we observe a coarser assignment resolution, but by slightly weighting towards their EDS signal we are able to improve the label assignment on several metrics.

Two approaches have been presented for selection of the number of principal components to retain for an AOI. Counting the number of grains in an EBSD quality map, for example with a Watershed algorithm, may be susceptible to oversampling. This can be due to the fact that coherent and chemically similar boundary precipitates will be counted separately but really should share a label. Slight oversampling is not a problem, but significant oversampling may make the components recovered after the VARIMAX rotation difficult to interpret. Furthermore it may reduce the ability of the method to amplify weak signals. A more systematic approach for selecting $n$ is to consider the contributions of retained principal components to total dataset variance and impose a limit beyond which we discard residual information as noise. A small $t$ (eg. 0.02%) corresponds to retaining many principal components as even those contributing relatively little signal are permitted to participate in the analysis. Increasing $t$ restricts the number of principal components, as we impose a low pass filter on the proportional variance contribution to the dataset required. This improves the signal-to-noise ratio in the RC-EBSPs and RC-spectra. We observe an increase in the mean and reduction in standard deviation of correlogram peak heights for cross correlation of measured and characteristic EBSD patterns and EDS spectra. When $t$ gets too high, the mean and standard deviation of correlation peak heights falls as we undersample the number principal components required to segment an AOI.

Optimal choice of parameters will depend on what dataset insight is required from an analysis. If we wish to reduce a dataset to as few RCCs as possible then a fairly tight tolerance limit provides a mechanism for quantitatively limiting the significance requirement of features in an AOI. If precipitates/grains of interest are small, reducing the tolerance limit permits weaker dataset signals to be assigned their own component. This may lead to oversampling of the dataset, reducing signal to noise.

Weighting PCA in favour of EBSD yields a finer effective spatial resolution in assignment than EDS weighting. This is advantageous if spatial precision is required, and especially when analysing phase presence in an AOI. Leveraging towards EDS can improve RCC assignment by several metrics. In some cases assigning chemically similar but structurally different regions (due to overlap in EDS interaction volume) to the same label may be compensated for by relaxing the variance tolerance limit and encouraging dataset oversampling. It may be the case that an EDS-weighted PCA would prove useful in situations where crystal pseudosymmetry or other similarity in Kikuchi bands reduces EBSD pattern contrast between two regions.

(a) Quantified RC-spectra

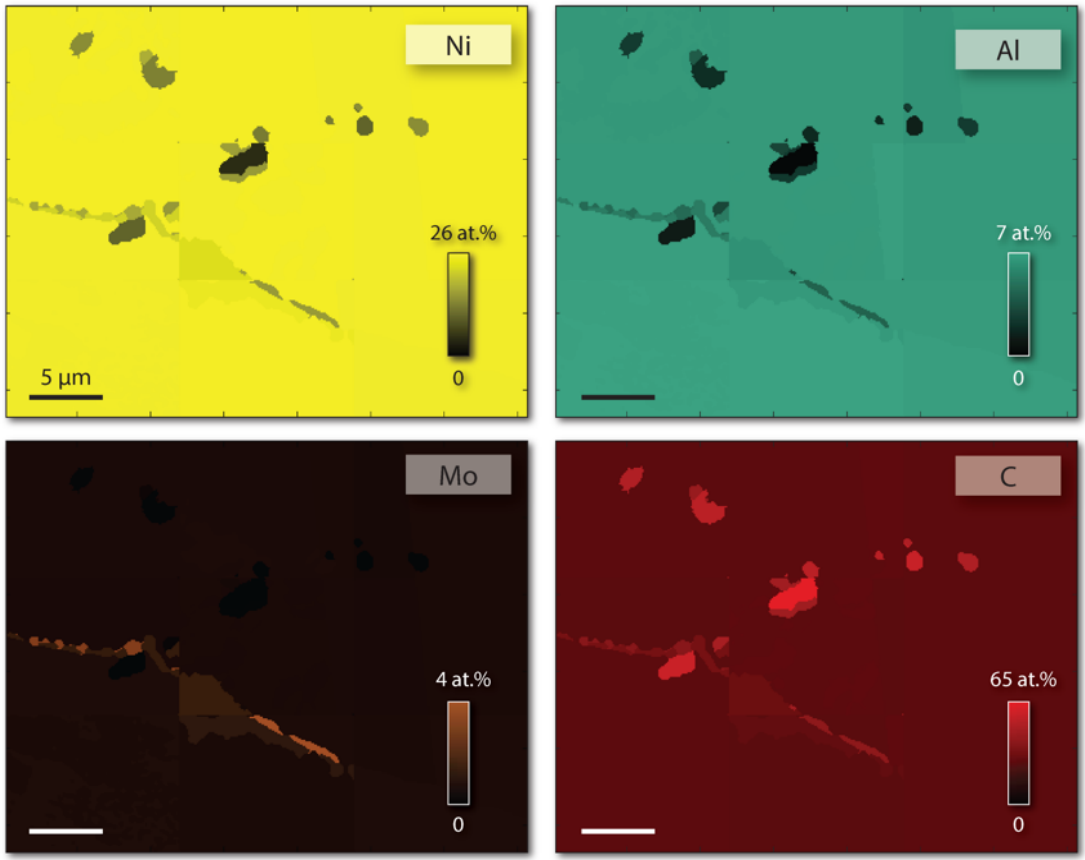

(b) Quantified average measured spectra

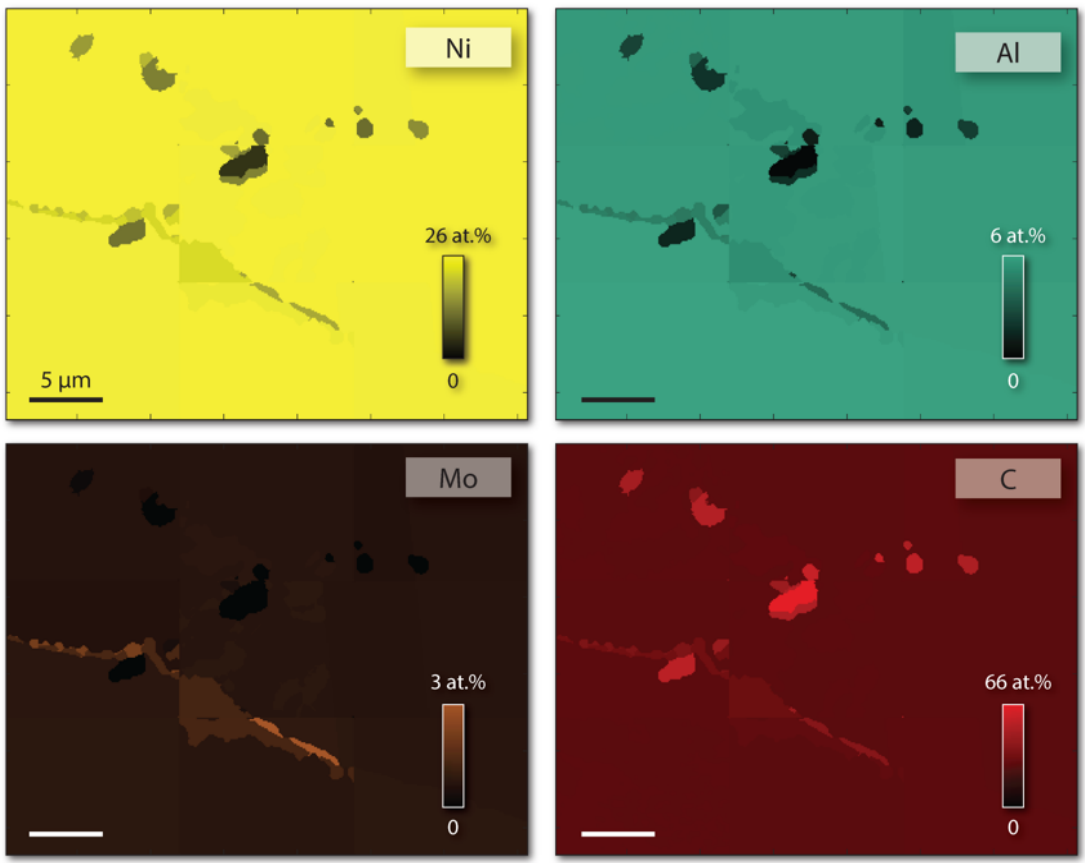

Figure 11: Chemical maps (at.%) of quantified RC-spectra. Only Ni, Al, Mo and C are shown for brevity. This analysis was performed with variance tolerance, $t$, of 0.2% and EBSD weighting, $w$, of 1. Maps are shown for directly quantified RC-spectra (a) and average spectra assigned to the same given label (b).

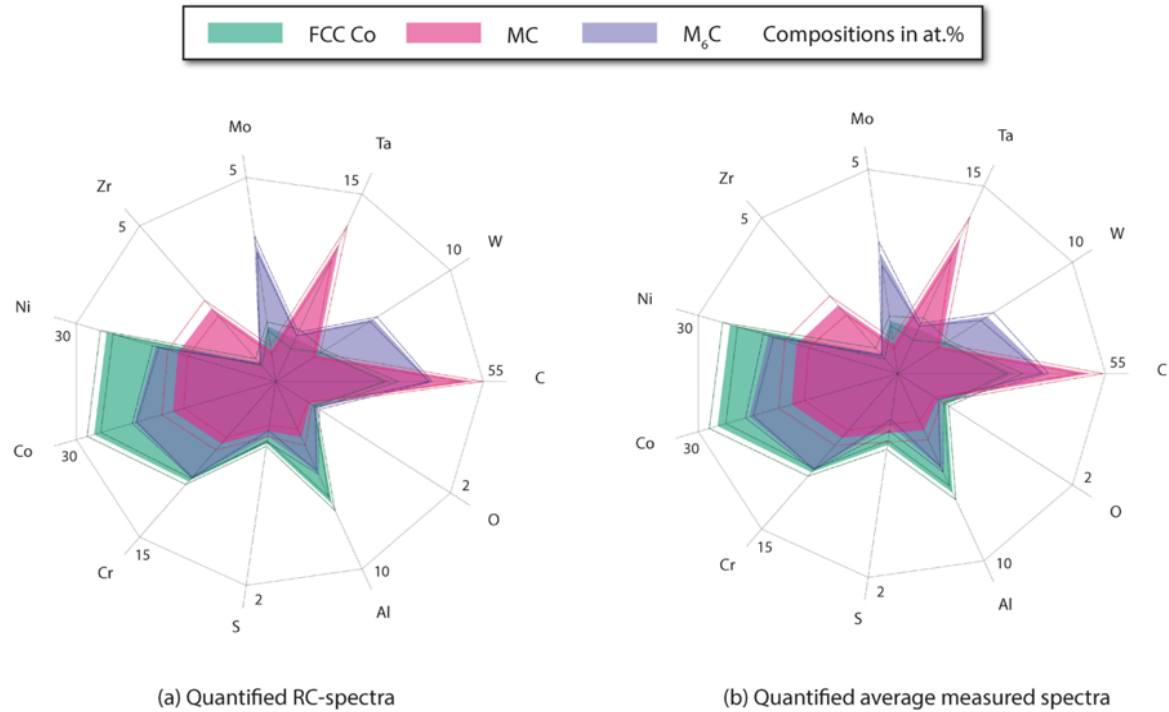

(a) Quantified RC-spectra

(b) Quantified average measured spectra

Figure 12: Average chemistry of the three identified phases in the dataset presented through this work. As with Figure 9a,c,e,g; Figure 10 and Figure 11, this analysis was performed with a variance tolerance, $t$, of 0.2% and an EBSD weighting, $w$, of 1. Filled regions plot average composition, dotted lines show +/- a standard deviation from the mean (tabulated data for this is provided in the supplementary information). This is shown for directly quantified RC-spectra (a) and average spectra assigned to the same given label (b).

### 4.2 Chemical analysis of labelled phases

Label chemistry can be quantified from RC-spectra independently of structure-ID from RC-EBSPs. Comparisons between chemistry and crystallography may then be made, with the benefit of a reduced signal to noise ratio of RC-spectra and RC-EBSPs than the raw measurements. RC-EBSPs and RC-spectra are simultaneously calculated and assigned to regions of an AOI. They are not independent, and reflect the most significant structural and chemical signals of points that they strongly load.

Figure 11 presents chemical maps (quantified RC-spectra) of a dataset for both directly quantified RC-spectra (a) and average spectra for a given RCC label (b). The same PCA parameters were employed as the results shown in Figure 9(a,c,e,g). They are almost identical. We observe that all precipitates exhibit Ni and Al depletion. The intergranular $M_6C$ carbides (Figure 9a) show Mo enrichment, while the intragranular MC carbides exhibit Mo depletion.

In Figure 12 we show elemental quantification of chemistry for each phase, along with standard deviations. This is performed for both directly quantified RC-spectra (11a) and quantified average spectra for a given label (11b). We note the trend in refractory element segregation between the carbides. In this system the MC carbide is strongly Ta and Zr enriched. Mo, Cr and W segregate to the $M_6C$ phase.

Similar observations of Ta and Zr enrichment in superalloy MC carbides has been reported by atom probe tomography studies [7,10]. We believe that the technique presented in this work provides a means of confirming trends in precipitate composition between alloys, as well as where elements tend to segregate upon nominal enrichment of the bulk composition.

### 4.3 Qualitative comparison to state-of-the art post-processing approaches

Our PCA approach is a method to amplify signal to noise, where we remove any knowledge of the co-location of measurement points and sum similar signals. In the limit, the EBSD pattern based neighbour pattern average method (NPAR) [41] improves signal to noise via a summation patterns within a local neighbourhood. This ignores extraction of the similarity of the signal obtained from each neighbouring pattern and could lead to adding of signals from two phases or grains which can affect interpretation of the average signal. The non-local pattern averaging reindexing (NLPAR) approach [42] provides delocalised smoothing, using a weighting function based upon the similarity of the pattern information in a moving window centred around a candidate point. Similar approaches have been adopted in the TEM community, particularly using the Hyperspy Python package, to obtain characteristic spot diffraction patterns for example with a cluster analysis approach [43]. So-called 'cluster-centre' diffraction patterns are calculated by grouping and calculating the average of DPs transformed into variable space. Both cluster-centre analysis and PCA can be considered extreme cases of NLPAR, where the spatial location is discarded and instead only applied statistics are relied on to denoise, label, and index data. Einsle *et al* [43] note that raw PCA is not suitable for spot diffraction analysis due to strong similarity in many reflections observed across the area of interest.

Generally, the challenge with a PCA approach is that the components returned represent the statistical dominance of each characteristic signal within the data set, and they are not physical. A VARIMAX rotation for the combined EBSD pattern and EDS spectra results in an easier to interpret label, where each label can be uniquely applied to each point within the map. This works for an EBSD pattern, as the variance between two Kikuchi-based diffraction patterns for different phases is relatively small. For a TEM spot-based diffraction pattern, rotating the data according to a variance model may not be reasonable, because the spot patterns for different phases may have a stronger variation in variance (e.g. due to a different number of reflectors that create spots within the pattern). In practice, this may impact how the diffraction data is pre-processed before putting into the data matrix, as well as a selection of an appropriate weighting scheme when joining the diffraction-based structure data with the EDS-based chemical data.

Keenan and Kotula [44,45] have explored scaling for multivariate statistical analysis for time of flight secondary ion mass spectrometry (TOF-SIMS) data. They focussed on how count-dependent pre-scaling impacts the distribution of variance and show that many scaling methods hinder multivariate statistical analysis. Their work highlights that variance scaling can be improved when the Poisson probability distribution of the raw data is accounted for, especially with regard to chemically significant minor features in the TOF-SIMS spectra. In contrast, our work combines two data types with different noise and scaling methods, and we use the output signals in separate characterisation processes.

When we consider the EDS signal, the heteroscedasticity of noise/error in the measurements is not just a function of the magnitude of the peak (which could be rectified by dividing by a Poisson correction factor for each energy bin, independent of the bin's energy) but also uncertainty in beam energy, sample elemental fluorescence and absorption, *etc.* Accounting for this could lead to formal normalisation of expected variance for a given energy bin, and would be a function of spectrum bin energy, identity of the chemical species, and intensity. Our method works well for alloy EDS datasets as the X-ray spectra contain many interacting (and covarying) signals. In a dataset there may for example be a majority of scan points with intense Ni and weak Mo peaks, producing a principal component that self-same pattern of intensity (intense Ni and weak Mo). An improved noise function would be useful where the EDS signal is less distinctly clustered. However, we note that in our workflow the final phase classification is performed on the EBSD signal, with the EDS data used for chemical quantification (interpreted as a function of structure). A similar but different noise model could be applied to each pixel within the EBSD pattern, dependent on the anisotropic spatial distribution of (near elastic) backscatter electrons that can be modulated in counts by Kikuchi diffraction. For example, this could be implemented through inclusion of a pixel and energy bin specific weighting function prior to operation of the 'macro' weighting term we have applied to mix the signals prior to applying the PCA that we have fairly extensively discussed.

In the absence of advanced noise models, we have employed a simpler method designed with our data processing stream in mind. This is focussed on (1) maximisation of the likelihood of successful segmentation of similar domains and the generation of appropriate characteristic signals; (2) the ability to register those domains against the EBSD signal for phase classification (which is augmented by the rotation of components to create uniform variance, as per Wilkinson *et al* [13]); (3) the subsequent EDS chemical signature analysis. Our approach is simple and sufficiently successful, and will be made available open source *via* AstroEBSD [27].

Improvements to our methodology could be motivated by the discussed prior work of Keenan and Kotula [44], provided noise and variance models of the EBSD signal and EDS signal can be determined and validated. At present, these are limited and the origins of the signals are still somewhat disputed [11,16,19,40]. While we have a significant grasp of the EBSD signal we are limited in our analysis of EDS information, and currently use standard proprietary software such as Bruker eSprit 2.1. Further method development could involve implementing open source quantification (accounting for discussed effects), including better accounting for Poisson noise within our processing toolbox.

## 5. Conclusions

We develop an analysis pipeline to provide robust correlative microscopy, mixing chemical information obtained using EDS and structural information obtained using EBSD. This enables us to observe small carbides and optimise signal-to-noise for the different phases present. PCA is an effective data processing technique for identifying regions of strong similarity in a dataset (microstructure), while remaining spatially unbiased (two adjacent points share the same propensity to be assigned the same principal component and RCC as two far-field points). Inclusion of both EBSPs and EDS spectra into the data matrix, D, provides a mechanism for obtaining simultaneous structural and chemical fingerprints of features in an AOI. It is possible to weight the identification of these characteristic EBSPs and spectra (RC-EBSPs and RC-spectra) in favour of similarity in crystallography or chemistry between points. In this work we present the following observations:

1. Selection of the number of principal components to retain for VARIMAX rotation and subsequent analysis can be made less subjective by counting grains in a Radon transform EBSP quality map, for example by a Watershed transform, or by selecting a tolerance limit (low pass filter) for the proportional explained variance of the retained principal components. Oversampling reduces signal to noise ratio in RCCs but reduces the risk of missing fine precipitates from the analysis.

2. An EBSD weighted PCA exhibits a finer effective spatial resolution in label assignment due to the smaller interaction volume of backscattered than secondary electrons, and therefore for EBSP than EDS-spectra generation.

3. Leveraging the PCA slightly towards EDS measurements can improve segmentation (lower standard deviation in cross-correlation peak height) of characteristic EBSPs and spectra.

4. Structural phase-ID of an AOI, for example by a Refined Template Matching algorithm, can be enhanced *via* data reduction of a 40,000 point map to (in the case of the dataset presented in this work) 35 RC-EBSPs. This drives a significant processing speedup, and permits the trialling of many candidate structure libraries, improving confidence in assignment.

5. Quantification of RC-spectra, assigned a structure label by RC-EBSP Refined Template Matching, permits measurements of average chemical segregation between phases.

## 6. Acknowledgements

TPM, TBB and DD would like to acknowledge support from the Rolls-Royce plc - EPSRC Strategic Partnership in Structural Metallic Systems for Gas Turbines (EP/M005607/1), and the Centre for Doctoral Training in Advanced Characterisation of Materials (EP/L015277/1) at Imperial College London. AF acknowledges funding from EPSRC. DDM and TBB acknowledge funding from the Shell AIMS UTC. CB and TBB acknowledge funding from the EPSRC Centre for Doctoral Training in Nuclear Engineering (EP/L015900/1) and Rolls-Royce plc. TBB would like to acknowledge the Royal Academy of Engineering for funding his research fellowship. DD acknowledges funding from the Royal Society for his industry fellowship with Rolls-Royce plc. We thank Angus Wilkinson (Oxford) for sharing MATLAB code used in [13] and for useful discussions. We used the Harvey Flower Microscopy Suite at Imperial College London for collecting this data. We would like to thank Mark Hardy (Rolls-Royce), Ioannis Bantounas, and Lucy Reynolds (both ICL), as well as Paraskevas Kontis and Baptiste Gault (both MPIE) for ongoing and very helpful discussions.

## 7. Author Contributions

TPM developed the majority of the MATLAB code in the work and drafted the initial manuscript. AF helped with adapting the refined template matching. DDM, AF, TBB, TPM, and CB discussed and developed ideas for the correlative approach. DD provided insight into the physical metallurgy of the alloy system and motivation for the study. CB performed CASINO electron interaction simulations for the phases identified by the analysis. DD and TBB supervised the work.

## 8. Data statement

Data will be uploaded to Zenodo upon article acceptance.